# Cancellation of RF Coupler-Induced Emittance Due to Astigmatism

David H. Dowell*
SLAC National Accelerator Laboratory
Stanford, CA 94309, USA

**Abstract**

It is well-known that the electron beam quality required for applications such as FEL's and ultra-fast electron diffraction can be degraded by the asymmetric fields introduced by the RF couplers of superconducting linacs. This effect is especially troublesome in the injector where the low energy beam from the gun is captured into the first high gradient accelerator section. Unfortunately modifying the established cavity design is expensive and time consuming, especially considering that only one or two sections are needed for an injector. Instead, it is important to analyze the coupler fields to understand their characteristics and help find less costly solutions for their cancellation and mitigation. This paper finds the RF coupler-induced emittance for short bunches is mostly due to the transverse spatial sloping or tilt of the field, rather than the field's time-dependence. It is shown that the distorting effects of the coupler can be canceled with a static (DC) quadrupole lens rotated about the z-axis.

**Introduction**

An important challenge for x-ray free electron lasers as light source facilities is increasing the number of simultaneous users. Currently these expensive facilities provide light to only one user at a time. One option is have a continuous pulse train rather than the currently used low repetition rate, short pulse trains. With the FEL operating at 100% duty factor (cw) it can be rapidly switched between nearly-simultaneous users. This path is already being taken by the superconducting RF (SRF) facilities at Flash and XFEL which operate at duty factors of a few percent, significantly above the $10^{-2}$ percent of LCLS-I. The LCLS-II XFEL now being designed will operate at 100% duty factor. However to succeed the newer cw machines will require solving a multitude of physics and engineering issues. This includes the minor but important topic of beam aberrations produced by the RF couplers feeding high power into the SRF cavities [1][2].

This paper investigates the field asymmetries of the power couplers of SRF linacs and how they distort and degrade the beam quality in the electron injector. Although these effects and corrections are reasonably well-understood, re-engineering established designs is usually too costly. Instead the lost FEL performance is recovered by using longer undulators or higher beam energy which are also expensive. Here we model the coupler effects on the beam, and investigate the means and methods for correcting and compensating its effects. This work attempts to first show how the coupler fields look like linear, tilted and rotated fields over the small region of the beam, and second to demonstrate compensation or cancellation of these fields with a quadrupole singlet rotated about the beam axis. Because the correcting rotated quadrupole's function is similar to that of multipole lenses used to correct aberrations in electron microscopes, one could call the rotated quadrupole a 'quadrupole stigmator' or in the jargon of electron microscopy: a 'four-pole stigmator'. A stigmator removes the xy correlations of an astigmatic beam.

*Corresponding email address: dowell@slac.stanford.edu

**Emittance due to RF coupler kicks**

The coupler's effect on the beam can be described as an instantaneous kick in voltage along the x, y and z-directions. Each component of the voltage kick consists of an instantaneous jump in voltage and transverse voltage gradient when the beam transits the coupler. Following the literature, the complex voltage kick factor is defined as [3]

$$\vec{v}(x,y) = \frac{\vec{V}(x,y)}{\vec{e_z}\cdot\vec{V}(0,0)} \cong \begin{pmatrix} v_{x0} + v_{xx}x + v_{xy}y \\ v_{y0} + v_{yx}x + v_{yy}y \\ 1 + \cdots \end{pmatrix} \tag{1}$$

The complex voltage kick, $\vec{V}(x,y)$, is given by integrals of the coupler fields along lines parallel to the z-axis (beam's optical axis),

$$\vec{V}(x,y) = \int [\vec{E}(\vec{r}) + c\vec{\beta} \times \vec{B}(\vec{r})]e^{i\omega z/c} dz \tag{2}$$

The $\vec{B}$-term is assumed to be imaginary to account for its $\pi/2$ RF phase shift with respect to the electric field. The complex voltage kick factor gives the electrons a momentum impulse of [3]

$$\vec{p} = Re\{\vec{v}(x,y)e^{i\phi_s}\}\frac{eV_{acc}}{c} \tag{3}$$

Here electron a distance $s$ behind the head electron has the phase $\phi_s = \frac{\omega s}{c} + \phi_{head}$ with respect to the linac RF. Writing out the components of the coupler's x-y plane momentum kick shows the spatial and phase dependences are separable into a complex voltage amplitude and phase, of which the real part is the momentum kick by the coupler,

$$\begin{pmatrix} p_x \\ p_y \end{pmatrix}_{coupler} = \frac{eV_{acc}}{c} Re\left\{\begin{pmatrix} v_{0x} + v_{xx}x + v_{xy}y \\ v_{0y} + v_{yx}x + v_{yy}y \end{pmatrix} e^{i\phi_s}\right\} \tag{4}$$

Dividing by the total momentum converts this to the coupler's angle kick,

$$\begin{pmatrix} x' \\ y' \end{pmatrix}_{coupler} = \frac{eV_{acc}}{\beta\gamma mc^2} Re\left\{\begin{pmatrix} v_{0x} + v_{xx}x + v_{xy}y \\ v_{0y} + v_{yx}x + v_{yy}y \end{pmatrix} e^{i\phi_s}\right\} \tag{5}$$

For simplicity, let us assume the electron bunch distribution in the xy-plane is uniform inside a square area having dimensions, $-R < x < R$ by $-R < y < R$. This trivial distribution simplifies the calculation of the variances and averages needed for deriving the coupler-induced emittance, while maintaining the essential physics. The variance and standard deviation for this *2R x 2R* square uniform distribution is,

$$\langle x^2 \rangle = \langle y^2 \rangle = \frac{R^2}{3} \quad \text{and} \quad \sigma_x = \frac{R}{\sqrt{3}} \tag{6}$$

The normalized emittance for the x-plane is defined as

$$\epsilon_n = \frac{\sqrt{\langle x^2\rangle\langle p_x^2\rangle - \langle xp_x\rangle^2}}{mc} = \sqrt{\langle x^2\rangle\langle x'^2\rangle - \langle xx'\rangle^2} \tag{7}$$

After some tedious algebra, the coupler-induced emittance is found to be solely due to the cross-term of the voltage kick $v_{xy}$,

$$\epsilon_{n,coupler} = \frac{eV_{acc}}{mc^2} \sigma_x^2 \left| v_{xy}^r \cos\left(\frac{\omega s}{c} + \phi_{head}\right) + v_{xy}^i \sin\left(\frac{\omega s}{c} + \phi_{head}\right) \right| \quad (8)$$

Here $v_{xy}^r$ and $v_{xy}^i$ are the real and imaginary parts of $v_{xy}$, respectively. This expression gives the emittance of a bunch slice the distance $s$ behind the bunch head. The RF phase of the bunch head is $\phi_{head}$ and the tail arrives after a bunch length, $s_{tail}$, and later at phase $\frac{\omega s_{tail}}{c} + \phi_{head}$. Figure 1 shows as an example the head and tail emittances vs. the RF phase for a head-tail phase difference of 10 degRF. It can be seen that the head and tail emittances are only slightly different; confirming the effect is mostly due to the non-linear transverse field rather than the phase-dependent kick. I.e., the emittance comes from the transverse spatial distribution of the coupler fields and not their time variation.

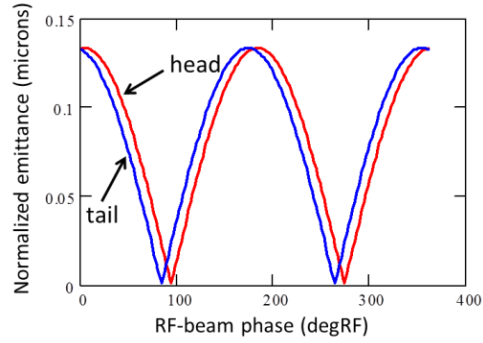

Figure 1: The head (s=0, red) and tail ($\frac{\omega s_{tail}}{c}$ =10degRF, blue) emittances vs. phase for $\sigma_x$ =1 mm; $V_{acc}$=20 MV and $v_{xy} = (3.4 + 0.2i)/mm$ [3].

**Modeling the RF coupler as a rotated quadrupole**

With the above definitions the vector equation for the transverse kick given to the beam by the coupler can be written more concisely as,

$$\begin{pmatrix} x' \\ y' \end{pmatrix}_{coupler} = \begin{pmatrix} \tilde{v}_{0x} + \tilde{v}_{xx}x + \tilde{v}_{xy}y \\ \tilde{v}_{0y} + \tilde{v}_{yx}x + \tilde{v}_{yy}y \end{pmatrix} \quad (9)$$

Here the real quantity $\tilde{v}$ is defined in terms of the complex voltage kick factors of Eqn. (1) and (2) and the slice phase with respect to the RF,

$$\tilde{v} \equiv \frac{eV_{acc}}{\beta\gamma mc^2} Re(ve^{i\phi_s}) \quad (10)$$

where $v = v_{0x}, v_{xx}, v_{xy}$, etc.

At this point we make the unwarranted assumption that

$$\tilde{v}_{yy} = -\tilde{v}_{xx} \quad \text{and} \quad \tilde{v}_{yx} = \tilde{v}_{xy} \quad (11)$$

These simple relations between the x- and y-plane kick factors result from Gauss' and Faraday's laws as given in Maxwell's equations. (A proof for Eqn. (11) can be found in the Appendix.) The validity of this symmetry between kick factors is evident in the Table 1 of Dohlus' paper which is reproduced here to illustrate this. The same symmetry can be seen in the analysis of the Cornell and LCLS-II couplers [4]. Therefore the discussion here will assume the Eqn. (11) relations are exactly valid and use them to re-express Eqn. (9) as

$$\begin{pmatrix} x' \\ y' \end{pmatrix}_{coupler} = \begin{pmatrix} \tilde{v}_{0x} + \tilde{v}_{xx}x + \tilde{v}_{xy}y \\ \tilde{v}_{0y} + \tilde{v}_{xy}x - \tilde{v}_{xx}y \end{pmatrix} \quad (12)$$

The $\tilde{v}_{0x,0y}$ terms are dipole kicks the coupler gives to the beam which can be cancelled with nearby steering dipoles. In simulations, this dipole steering is seen to be very effective at mitigating the dipole-kick emittance. Similarly a quadrupole field with the proper rotation and strength can correct for the coupler's lens-like kicks. Since the $\tilde{v}_{xx}$ and $\tilde{v}_{yy}$ terms are linear with x and y, resp. they effect the beam like a focusing or defocusing lens. This focusing is linear and by itself does not produce any emittance. Instead it is the cross terms, $\tilde{v}_{xy}$ and $\tilde{v}_{yx}$, are due to a correlation between x and y which is the principal cause of emittance.

Table 1: RF kick coefficients

| | upstream | downstream |
|---|---|---|
| $v_{x0} \cdot 10^6$ | -57+7i | -23+52i |
| $v_{xx} \cdot 10^6$/mm | 1.0-0.7i | -3.7-2i |
| $v_{xy} \cdot 10^6$/mm | 3.4+0.2i | 3.0+0.4i |
| $v_{y0} \cdot 10^6$ | -42-3i | 30+5i |
| $v_{yx} \cdot 10^6$/mm | 3.4+0.2i | 3.0+0.4i |
| $v_{yy} \cdot 10^6$/mm | -1.1+0.6i | 3.8+1.9i |

Table reproduced from Ref [3] with added highlights (color).

Eqn. (12) shows why the above assumption was chosen. With this symmetry one can easily see the angle of the kick in the x-y plane can be represented by the rotation of a constant quadrupole field about the z-axis. Thus the $v_{xy}y$ term of the x-component comes from the axial rotation of the coupler which correlates the x- and y-plane components. This correlation appears as uncorrelated emittance in the x- and y-plane trace-space emittances. This is much like the relation between the slice and projected emittances. There the longitudinal correlation between slices appears as an uncorrelated projected emittance.

Generally uncorrected correlations can be problematic since if the correlation becomes 'diluted' or random, then the correlations is lost and even the 4-D canonical emittance will increase. However having the correlation also means that compensation is possible and the coupler-induced emittance eliminated. The next section describes how the SRF coupler field effects can be cancelled with a rotated quadrupole field which compensates for the coupler-induced correlation. In addition, because the offending coupler field asymmetry is due to the rotation of a constant quadrupole field, a simple DC quadrupole having the proper rotation angle and field strength can be used for the correction. This approach is discussed next.

**Compensation of coupler kicks with a rotated quadrupole**

As previously discussed, a thin quadrupole lens rotated an angle $\theta_q$ about the z-axis gives the beam an angular kick very similar to the kick given by the coupler. Therefore we explore using a skewed quadrupole whose rotation angle and focal length can be adjusted to undo the coupler rotation and cancel its focusing strength, thereby eliminating its effect on beam emittance. It can be shown that the kick angle vector in the xy-plane for a quadrupole with focal length $f_q$ and rotated $\theta_q$ about the z-axis summed with x- and y-dipole kicks can be written as

$$\begin{pmatrix} x' \\ y' \end{pmatrix}_{dipole+quad} = \begin{pmatrix} x_0' - \frac{\cos 2\theta_q}{f_q} x - \frac{\sin 2\theta_q}{f_q} y \\ y_0' - \frac{\sin 2\theta_q}{f_q} x + \frac{\cos 2\theta_q}{f_q} y \end{pmatrix} \tag{13}$$

Here it is useful to point out the similarities between the coupler kicks and those of a rotated quad. By associating terms between matrices in Eqns. (12) and (13) one finds that,

$$\tilde{v}_{xx} = -\frac{\cos 2\theta_c}{f_c} \quad \text{and} \quad \tilde{v}_{xy} = \frac{\sin 2\theta_c}{f_c} \tag{14}$$

These relations allow us to model the coupler fields equivalently as a quadrupole with focal length $f_c$ rotated $\theta_c$ about the z-axis. The equivalent quadrupole rotation angle of the coupler field is then

$$\theta_c = -\frac{1}{2} \tan^{-1} \frac{\tilde{v}_{xy}}{\tilde{v}_{xx}} \tag{15}$$

Summarizing results so far, it's been determined that the basic behavior of the coupler field can be modeled as a thin, rotated quadrupole with focal length $f_c$ and rotation angle $\theta_c$. The $x_0'$ and $y_0'$ kicks are included in Eqn. (13) for completeness to explicitly show the need for dipole correctors as well as for the rotated quadrupole currently being discussed. Generally it is assumed that nearby steering is available to cancel the dipole kicks such that $x_0' = -\tilde{v}_{0x}$, and similarly for y. Therefore these kick terms are ignored in what follows. And finally the analysis will assume the quadrupole and coupler fields are thin compared to their focal lengths and located close to each other on the beamline. Then Eqns. (12) and (13) (without the dipole kicks) can be added to give the total angle kick due to the quadrupole fields,

$$\begin{pmatrix} x' \\ y' \end{pmatrix}_{total} = \begin{pmatrix} x' \\ y' \end{pmatrix}_{coupler} + \begin{pmatrix} x' \\ y' \end{pmatrix}_{quad} = \begin{pmatrix} \left\{\tilde{v}_{xx} - \frac{\cos 2\theta_q}{f_q}\right\} x + \left\{\tilde{v}_{xy} - \frac{\sin 2\theta_q}{f_q}\right\} y \\ \left\{\tilde{v}_{xy} - \frac{\sin 2\theta_q}{f_q}\right\} x - \left\{\tilde{v}_{xy} - \frac{\cos 2\theta_q}{f_q}\right\} y \end{pmatrix} \tag{16}$$

The beauty of the Eqn. (11) symmetry can now be appreciated in this equation. It shows that both the emittance and the focusing effects of the coupler can be eliminated by solving for the quadrupole rotation and focal strength which make the coefficients of all x- and y-terms zero. Given the symmetry of the expression, this occurs when the following two (rather than four) conditions are satisfied,

$$\tilde{v}_{xx} - \frac{\cos 2\tilde{\theta}_q}{\tilde{f}_q} = 0 \quad \text{and} \quad \tilde{v}_{xy} - \frac{\sin 2\tilde{\theta}_q}{\tilde{f}_q} = 0 \tag{17}$$

Simultaneous solutions of these two equations gives paired values for the quad's rotation angle and focal strength which cancel the coupler field's cross-term (emittance) and the quadrupole-focus term (astigmatism). Doing this yields the solutions for these quadrupole parameters in terms of the coupler's complex voltage kick,

$$\tilde{\theta}_q = \frac{1}{2} \tan^{-1} \frac{\tilde{v}_{xy}}{\tilde{v}_{xx}} \tag{18}$$

$$\frac{1}{\tilde{f}_q} = \frac{eV_{acc}}{\beta\gamma mc^2} \sqrt{\tilde{v}_{xx}^2 + \tilde{v}_{xy}^2} \tag{19}$$

The normalized voltage kick factors, $\tilde{v}$, in terms of the accelerator voltage and slice phase are,

$$\tilde{v}_{xx}(\phi_s) = \frac{eV_{acc}}{\beta\gamma mc^2}\left(v_{xx}^r \cos\phi_s - v_{xx}^i \sin\phi_s\right) \tag{20}$$

and

$$\tilde{v}_{xy}(\phi_s) = \frac{eV_{acc}}{\beta\gamma mc^2}\left(v_{xy}^r \cos\phi_s - v_{xy}^i \sin\phi_s\right) \tag{21}$$

Inserting these relations into Eqns. (18) and (19) gives the RF phase-dependent solutions for the quadrupole rotation angle,

$$\tilde{\theta}_q(\phi_s) = \frac{1}{2}\tan^{-1}\frac{v_{xy}^r \cos\phi_s - v_{xy}^i \sin\phi_s}{v_{xx}^r \cos\phi_s - v_{xx}^i \sin\phi_s} \tag{22}$$

and the quadrupole focal strength,

$$\frac{1}{\tilde{f}_{q(\phi_s)}} = \frac{eV_{acc}}{\beta\gamma mc^2}\sqrt{\left(v_{xx}^r \cos\phi_s - v_{xx}^i \sin\phi_s\right)^2 + \left(v_{xy}^r \cos\phi_s - v_{xy}^i \sin\phi_s\right)^2} \tag{23}$$

which cancel the coupler's kicks. Together these two expressions satisfy Eqns. (18) and (19), and thereby eliminate all first-order coupler-induced rotation and focusing effects on the beam.

As an example we again use the values in Dohlus' paper [3] to compute the lens requirements. His Table 1 gives the normalized complex voltage kick factors as

$$v_{xx} = (1 - 0.7i) \times 10^{-6} \text{ /mm} \tag{24a}$$
$$v_{xy} = (3.4 - 0.2i) \times 10^{-6} \text{ /mm} \tag{24b}$$

These coupler kicks correspond to a normalizing voltage of 20 MV. Assume a beam kinetic energy of 800 KeV such that $\beta\gamma = 2.5$. The quadrupole rotation angle (blue) and focal length (red) needed for the correction as functions of the bunch-head phase with respect to the RF are plotted in Figure 1 using Eqns. (22) and (23), respectively.

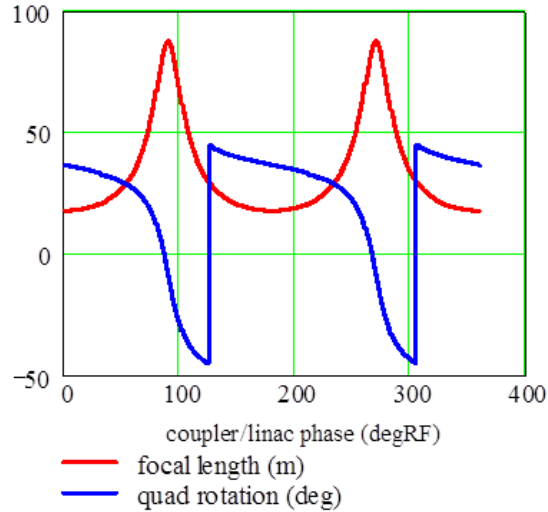

Figure 1: The correction quadrupole angle (blue) and focal length (red) vs. the coupler RF phase.

**Summary and conclusions**

This paper discussed the effects of field asymmetries of the power couplers of SRF linacs upon electron beams. For some applications, the field asymmetry of the coupler designs used in SRF linacs can significantly degrade the beam quality. However the current couplers work well for their primary purpose and it is too expensive to re-engineer and test new couplers for only a few linac structures.

Therefore it is important to more fully understand how the coupler affects the beam and investigate other means and methods for correcting and compensating its effects.

This work shows the coupler-induced emittance is dominantly due to a rotation about the z-axis of the field gradients and not by the time-dependence of the RF fields. Thus all slices get the same emittance kick due to the coupling of the transverse degrees of freedom. It is demonstrated that the coupler optics can be modeled as a rotated quadrupole with focal length and rotation angle given in terms of the complex voltage kicks. In addition the emittance is proved to be due to the skewed component of the coupler's quadrupole field, and that the emittance results principally from a xy-correlation due to the rotational kicks of the beam by the coupler field. The analysis shows that a rotated quadrupole located within a focal length of the coupler is effective at compensating for the coupler's kicks, cancelling both the coupler emittance and the astigmatic focusing. Expressions are given for the quadrupole rotation angle and focal strength in terms of the coupler's RF phase, peak field and effective length.

This 'quadrupole-stigmator' will generally consist of a pair of quadrupole singlets arranged such that one has its poles in normal alignment with the xy-plane and the second rotated to be a skewed (45 degree rotated) quadrupole. The rotation angle of the total quadrupole field results from the relative strengths of the two quads and the total strength is the vector sum.

And lastly it needs mentioning that the quadrupole stigmator has broader implications beyond just aberration correction of RF couplers. This is because the stigmator corrects for all skewed quadrupole fields within a section of beamline, no matter what makes them or where they are produced. In fact, an initial implementation of quadrupole correctors in a photoinjector compensated for an anomalous quadrupole field in a solenoid [5]. Therefore it is now becoming apparent that multipole correctors will play a central role in this next round of emittance reduction in photoinjectors.

**Acknowledgements**

I wish to acknowledge stimulating discussions with my SLAC, LBNL and Cornell U. colleagues on the topics of coupler aberration and compensation. I especially wish to express my gratitude to John Schmerge for discussions and support on the LCLS-II project.

## Appendix

Prove that if the transverse electric field over a small region near the beam axis is specified as a linear expansion

$$E_x = E_{x,0} + \frac{\partial E_x}{\partial x}x + \frac{\partial E_x}{\partial y}y \tag{A1a}$$

$$E_y = E_{y,0} + \frac{\partial E_y}{\partial x}x + \frac{\partial E_y}{\partial y}y \tag{A1b}$$

then the field gradients are related as

$$\frac{\partial E_y}{\partial y} = -\frac{\partial E_x}{\partial x} \quad \text{and} \quad \frac{\partial E_y}{\partial x} = \frac{\partial E_x}{\partial y} \,. \tag{A2}$$

*Proof:*

Gauss's law gives the divergence of the electric field for no beam charge and only the field, $\vec{E}$, as

$$\vec{\nabla} \cdot \vec{E} = \frac{\partial E_x}{\partial x} + \frac{\partial E_y}{\partial y} + \frac{\partial E_z}{\partial z} = 0 \tag{A3}$$

For the case of the coupler fields, assume the longitudinal field varies slowly with z compared to the gradients of the transverse fields. Thus the $\frac{\partial E_z}{\partial z}$ term is zero and we thereby obtain the first relation in Eqn. (A2),

$$\frac{\partial E_x}{\partial x} + \frac{\partial E_y}{\partial y} = 0 \quad => \quad \frac{\partial E_y}{\partial y} = -\frac{\partial E_x}{\partial x} \tag{A4}$$

Therefore Gauss's law leads to the diagonal derivatives of $\vec{E}$ being anti-symmetric (equal with opposite signs).

Next consider Faraday's law to determine the symmetry of the cross-term voltage kicks like $v_{xy}$,

$$\vec{\nabla} \times \vec{E} = -\frac{\partial \vec{B}}{\partial t} \tag{A5}$$

Here the derivative on the right can be ignored for two reasons: 1) the bunch is short, near velocity c and thereby samples a frozen-in-time snapshot of the fields and 2) for a TM-mode field $B_z$=0. Since $\vec{E}$ does not depend upon z, only the z-term of the cross product is non-zero,

$$\frac{\partial E_y}{\partial x} - \frac{\partial E_x}{\partial y} = -\frac{\partial B_z}{\partial t} = 0 \tag{A6}$$

Leading to symmetric (equal) cross-gradients,

$$\frac{\partial E_y}{\partial x} = \frac{\partial E_x}{\partial y} \tag{A6}$$

Thus the second part of Eqn. (A2) is proved.

Therefore it has been shown that

$$\frac{\partial E_y}{\partial y} = -\frac{\partial E_x}{\partial x} \quad \text{and} \quad \frac{\partial E_y}{\partial x} = \frac{\partial E_x}{\partial y} \,. \tag{A7}$$

And within a small region near the beam axis the local field is given by

$$E_x = E_{x,0} + \frac{\partial E_x}{\partial x}x + \frac{\partial E_x}{\partial y}y \tag{A8a}$$

$$E_y = E_{y,0} + \frac{\partial E_x}{\partial y}x - \frac{\partial E_x}{\partial x}y \tag{A8b}$$

QED!